# Large and layer dependent nonlinear optical absorption of MXene 2D thin films

David J. Moss

Optical Sciences Centre, Swinburne University of Technology, Hawthorn, Vic. 3122, Australia

*E-mail: *dmoss@swin.edu.au*

**Abstract**

As a rapidly expanding family of two-dimensional (2D) materials, MXenes have recently gained considerable attention due to their appealing properties. Here, by developing a solution-based coating method that enables transfer-free and layer-by-layer film coating, we investigate the layer-dependent nonlinear optical absorption of $Ti_3C_2T_x$ films – an important member of the MXene family. By using the Z-scan technique, we characterize the nonlinear absorption of the prepared MXene films consisting of different numbers of monolayers. The results show that there is a strong and layer-dependent nonlinear absorption behavior, transitioning from revisable saturable absorption (RSA) to saturable absorption (SA) as the layer number increases from 5 to 30. Notably, the nonlinear absorption coefficient $\beta$ varies significantly within this range, changing from $\sim 7.13 \times 10^2$ cm/GW to $\sim -2.69 \times 10^2$ cm/GW. We also characterize the power-dependent nonlinear absorption of the MXene films at various incident laser intensities, and a decreasing trend in $\beta$ is observed for increasing laser intensity. These results reveal the intriguing layer-dependent nonlinear optical properties of 2D MXene films, highlighting their versatility and potential for implementing high-performance nonlinear photonic devices.

# 1. INTRODUCTION

Nonlinear photonic devices offer a powerful solution for realizing ultrafast information processing through all-optical signal processing, which surpasses the capabilities of electronic processing by providing speeds that are several orders of magnitude higher.[1-3] As fundamental building blocks for implementing nonlinear photonic devices, advanced optical materials with excellent nonlinear optical properties have been extensively investigated.[4, 5] Recently, there has been increasing interest in the nonlinear optical properties of two-dimensional (2D) materials,[6-11] which have atomically thin structures and exhibit many remarkable properties that are much superior to those of conventional bulk materials.[12-14] A variety of 2D materials, such as graphene,[15, 16] graphene oxide (GO),[17, 18] transition metal dichalcogenides (TMDCs),[8, 19, 20] black phosphorus (BP),[21-23] perovskite,[9, 24] and MXene,[25-27] have been investigated, exhibiting attractive properties such as strong saturable absorption (SA) or revisable saturable absorption (RSA), ultrahigh second or third-order optical nonlinearity, significant material anisotropy, and broadband response. These properties have enabled the development of various nonlinear photonic devices for diverse applications, such as mode-locking lasers,[28, 29] all-optical modulators,[30, 31] polarization-dependent all-optical switches,[22, 32] and nonlinear optical generation and processing.[33-37]

As a new category of 2D materials that has received significant attention in recent years, MXenes have shown many exceptional mechanical, thermal, electrical, and optical properties.[38, 39] For example, the rich surface groups and flexible layer spacing in MXene make it a highly effective photocatalyst.[40, 41] Moreover, MXenes have the capability to absorb near-infrared radiation, resulting in elevated photothermal conversion efficacy.[42, 43] The exceptional plasma characteristics have also underpinned the realization of plasma photodetectors and surface-enhanced Raman spectroscopy (SERS).[44, 45] Recently,[26, 27] it has been reported that MXenes exhibit significant nonlinear absorption that is two orders of magnitude higher than BP[46] and molybdenum disulfide ($MoS_2$).[47] However, no detailed characterization has been conducted to examine how the nonlinear absorption properties are influenced by the film thickness.

In this paper, we prepare layered $Ti_3C_2T_x$ films via a solution-based method that yields transfer-free and layer-by-layer film coating, which allows us to investigate the layer-dependent nonlinear absorption of the 2D MXene films that has not been characterized previously. We utilize the Z-scan technique to measure the nonlinear absorption of the MXene films, and the results reveal a strong layer-dependent nonlinear absorption behavior – transitioning from reversible saturable absorption (RSA) to saturable absorption (SA) as the layer number increases from 5 to 30. Remarkably, the nonlinear absorption coefficient $β$ undergoes a considerable change within this range, varying from ~7.13 × $10^2$ cm/GW to ~-2.69 × $10^2$ cm/GW. In addition, we characterize the nonlinear response of the MXene films at varying incident laser intensities and observe a decreasing trend in the nonlinear absorption coefficient $β$ as the laser intensity increases. These results reveal interesting insights about the evolution of the nonlinear optical properties of 2D MXene as the film thickness increases. Furthermore, our MXene film coating method is highly compatible with integrated photonic devices. All of these paves the way for their applications in high-performance nonlinear photonic devices.

## 2. MATERIAL PREPARATION AND CHARACTERIZATION

**Figure 1a** illustrates the atomic structure of $Ti_3C_2T_x$, which is an important member of the MXene family that has been widely studied.[48] The basic structure of $Ti_3C_2T_x$ is composed of alternating stacks of Ti and C atoms, and thick $Ti_3C_2T_x$ films are made up of $Ti_3C_2$ layers that are divided by groups containing randomly distributed elements such as -O, -OH, and -F. The surface functional groups of MXene materials play a crucial role in determining their properties.[49] For example, the hydroxylated and fluorinated terminations are more transparent, while the oxygen termination increases both absorption and reflectance at visible wavelengths.[50] In addition, upon absorbing light, the energy of the photon in MXene film could transform into lattice motion inside the atomic structure, leading to the generation of phonons as well as the increase of the film temperature.[43]

We fabricated MXene films on dielectric substrates through self-assembly of MXene nanoflakes in a solution by electrostatic attachment. **Figure 1b** illustrates the fabrication process flow. Before film coating, a MXene solution and a polymer solution were prepared. The former contained negatively charged 2D MXene nanoflakes synthesized through the LiF/HCl-etching method,[39, 51] whereas the latter contained positively charged polyelectrolyte polydiallyldimethylammonium chloride (PDDA) polymer. To construct multi-layered films on the target substrate, the process of depositing a single monolayer MXene film was repeated, which involved four steps. First, the silica substrate with a negatively charged surface was immersed in the prepared polymer solution, resulting in the formation of a polymer-coated substrate with a positively charged surface. Second, the polymer-coated substrate was rinsed with a stream of deionized distilled water and then dried using $N_2$. Third, the polymer-coated substrate was submerged in the prepared MXene solution, allowing for the formation of an MXene monolayer on the upper surface driven by electrostatic forces. Finally, the MXene-coated substrate was rinsed with a stream of deionized distilled water and then dried using $N_2$. Unlike the cumbersome transfer processes employed for coating other 2D materials such as graphene and TMDCs,[52, 53] our method enables transfer-free and layer-by-layer coating of MXene films on dielectric substrates, together with high scalability and accurate control of the layer number or the film thickness. This coating method also shows high compatibility with integrated devices. Previously, we used a similar method to coat 2D layered GO films and successfully demonstrated many functional integrated photonic devices.[17, 18, 33, 54-56]

**Figure 1c** shows a picture of the as-prepared MXene film coated on a silica substrate, where a high degree of homogeneity is observed over the entire substrate. According to atomic force microscopy (AFM) measurements, the average thickness of a monolayer MXene film is ~3 nm. **Figure 1d** shows the Raman spectra of the MXene films with different layer numbers $N$ = 5, 10, 15, 20, and 30, which were measured using a ~514-nm pump laser. For all the samples, two narrow peaks at around ~203 cm$^{-1}$ and ~734 cm$^{-1}$ were observed. As the layer number increases, the magnitude of the peaks also increases. These results are consistent with previous measurements for MXene films in Refs. [57] and [58] which verify the high quality of our prepared MXene films.

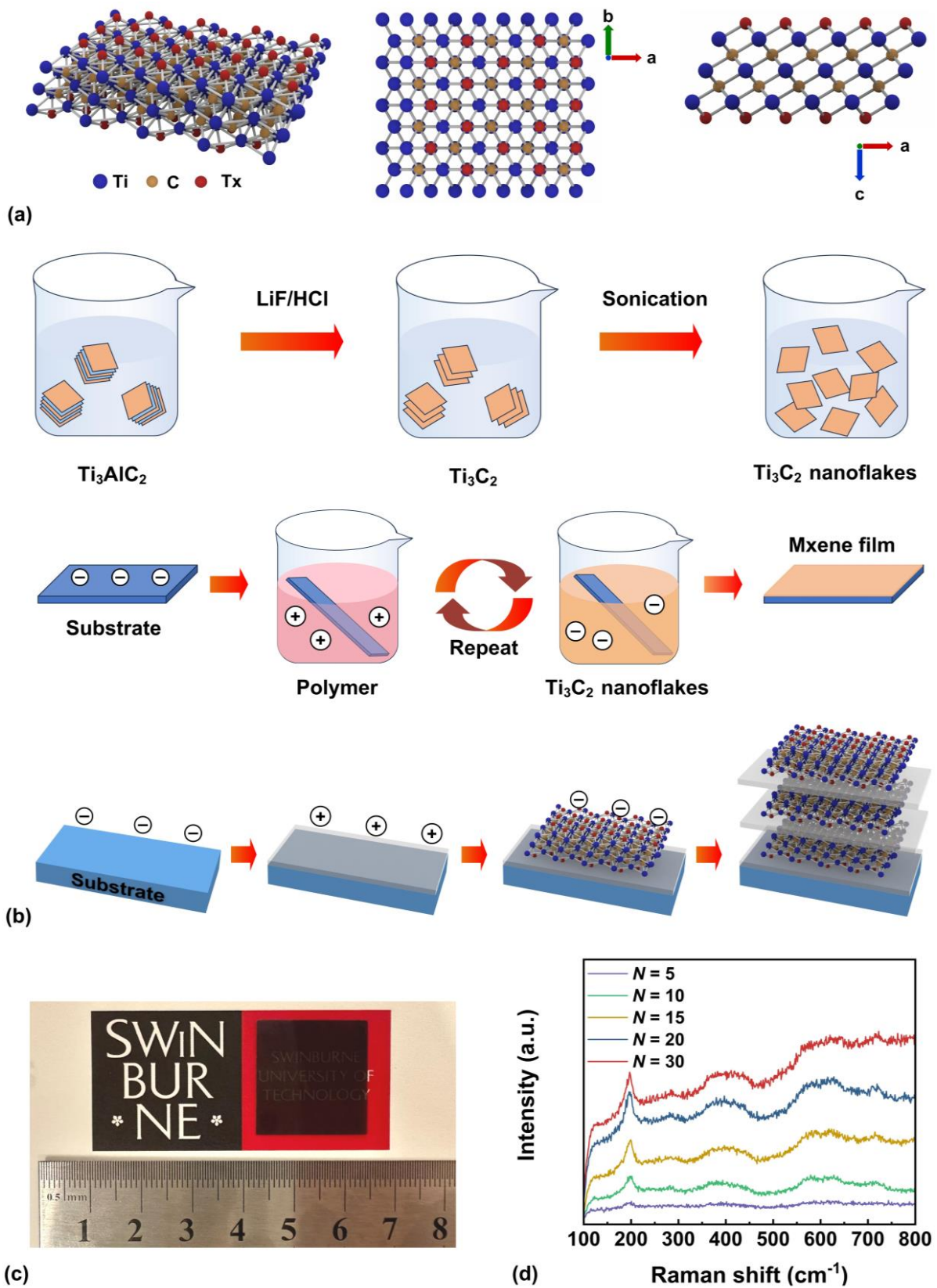

**Figure 1.** (a) Schematic atomic structure of MXene consisting of layered $Ti_3C_2$. (b) Schematic illustration of process flow used to fabricate MXene films via self-assembly. The top row depicts the process flow for preparing a solution containing 2D MXene nanoflakes. The middle row illustrates the process flow for coating the MXene film on a substrate through self-assembly. The bottom row illustrates the progression of the sample during the self-assembly process. (c) Photograph of a 10-layer MXene film coated on a silica substrate. The ruler unit is in centimeters. (d) Raman spectra for the prepared MXene films with different layer numbers $N$ = 5, 10, 15, 20, and 30.

**Figure 2a** shows the optical absorption spectra of the MXene films with different layer numbers $N$, which were characterized by ultraviolet–visible (UV–vis) spectrometry. The linear absorption spectrum when $N$ = 5 decreases sharply at wavelengths < 400 nm and exhibits a low absorption at wavelengths > 600 nm. In contrast, when $N$ > 5, the absorption spectra rise rapidly and then fall rapidly in the range of 300 – 500 nm. In addition, the linear absorption of the samples increases with the increase of layer number $N$.

The optical bandgap of the MXene film can be estimated from a Tauc plot of $(\alpha h v)^{1/2}$ versus $hv$ using the Tauc formula,[59] where $\alpha$ and $h$ are the optical absorption coefficient and photon energy, respectively. **Figure 2b** shows the Tauc plot extracted from the linear absorption spectra in **Figure 2a**, and **Figure 2c** shows the optical bandgap versus layer number further derived from **Figure 2b**. As the layer number $N$ increases from 5 to 30, the optical bandgap of the MXene films decreases from ∼1 eV to ∼0.66 eV. We also note that the bandgaps of our prepared MXene films are slightly higher than the values reported in the previous literature.[41] This is possibly due to the presence of titanium oxide on the surface of MXene.[60] Furthermore, different fractions of the surface functional groups could also result in the variations in the bandgap.[61]

**Figure 2d** shows the transmittance spectra of the MXene films with different layer numbers $N$ = 5, 10, 15, 20, and 30. The transmittance of the samples decreases with an increasing layer number. The 5-layer sample has a transmittance > 60% at wavelengths between 400 nm and 1800 nm, which is consistent with previous results in Refs. 26 and 62.

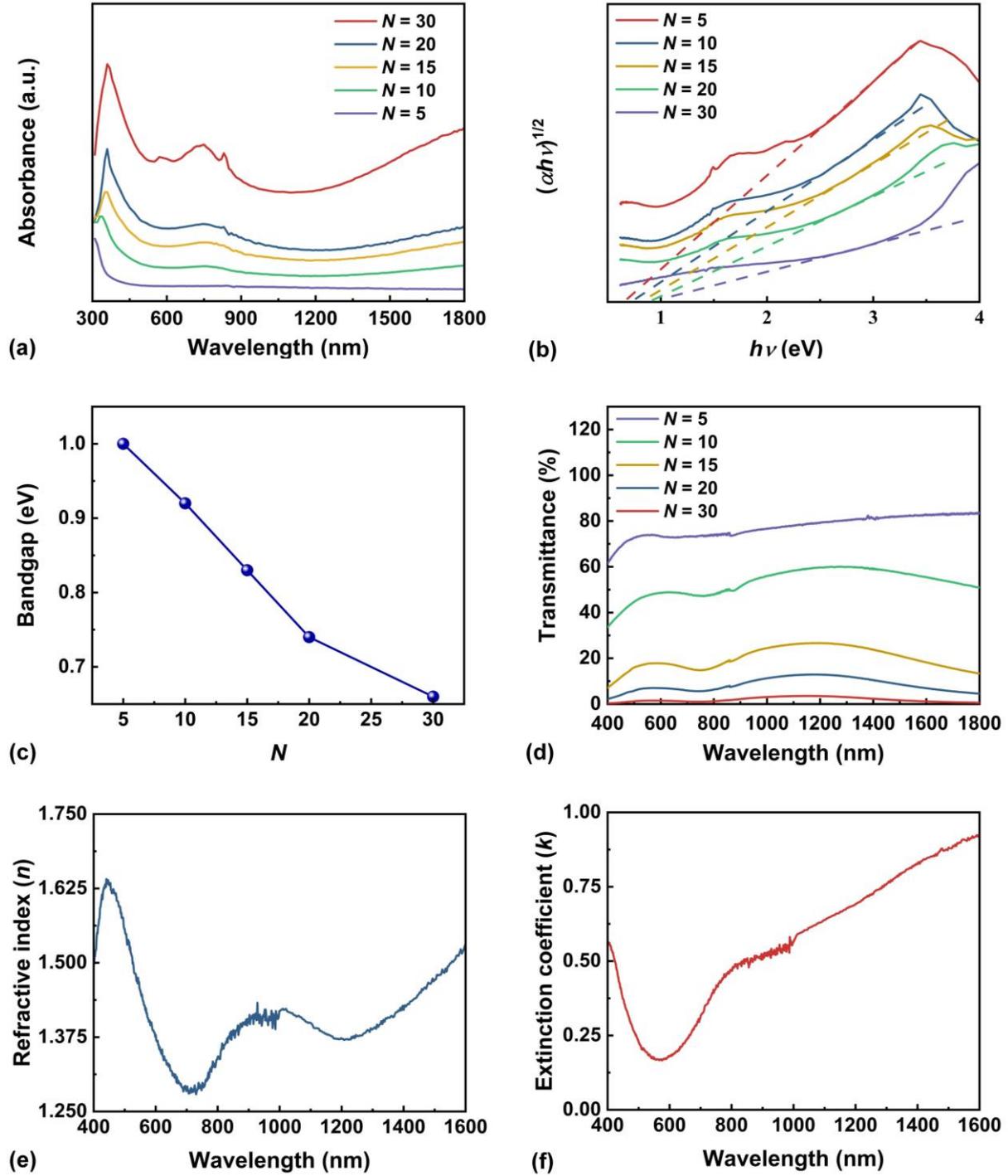

**Figure 2.** Characterization of the prepared MXene films. (a) UV-vis absorption spectra for samples with different layer numbers $N$. (b) The Tauc plot extracted from the absorption spectra in (a). (c) Optical bandgap versus $N$ extracted from (b). (d) Linear transmittance spectra for samples with different layer numbers $N$. (e)-(f) Measured in-plane refractive index $n$ and extinction coefficient $k$ for the 5-layer sample.

**Figures 2e** and **2f** show the in-plane refractive index $n$ and extinction coefficient $k$ of the 5-layer MXene film characterized by spectral ellipsometry, respectively. Since the out-of-plane response of the thin samples is much weaker, we could only measure the in-plane $n$ and $k$ of the

MXene film. The refractive index first increases sharply, reaching its peak at ~500 nm. It then experiences a sharp decline, reaching a minimum at ~700 nm, followed by a fluctuating rise. This shows an agreement with the trend of the UV−vis absorption spectra in **Figure 2a** and confirms the validity of our ellipsometry measurements. The trend for $k$ is opposite to that of $n$, with $k$ falling sharply and reaching a minimum at ~600 nm, and then increasing rapidly before slowing down at ~800nm. Our measured $k$ values are lower than that in Ref. 43, which can be attributed to the differences in the sample size and the surface functional groups of the MXene films. In addition to the 5-layer sample, we conducted measurements on other samples with different layer numbers. The measured values for $n$ were nearly identical to those obtained for the 5-layer sample. The measured $k$ values exhibit a minor rise as the layer number increases. For instance, at a wavelength of ~800 nm, the measured $k$ values for the samples including 5 and 30 layers were ~0.49 and ~0.53 respectively.

## 3. Z-SCAN MEASUREMENTS

The Z-scan technique was employed to characterize the nonlinear optical absorption of the MXene films that we prepared. **Figure 3a** illustrates the experimental setup used for the Z-scan measurements. The samples were excited using femtosecond optical pulses generated by an optical parametric oscillator, which had a center wavelength of ~800 nm, a repetition frequency of ~80 MHz, and a pulse duration of ~140 fs. The utilization of a half-wave plate in conjunction with a linear polarizer was implemented as a power attenuator to adjust the incident light power. The beam expansion system consisted of a 25-mm concave lens and two 150-mm convex lenses, which were utilized to expand the light beam. The expanded beam was then focused by an objective lens (10×, 0.25 NA), resulting in a focused spot size of ~1.6 μm. The prepared sample to be measured was positioned at a right angle to the direction of the beam axis and subsequently moved along the Z-axis using a highly precise one-dimensional (1D) linear motorized stage. The alignment of the light beam to the target sample was achieved through a high-definition charge-coupled device (CCD) imaging system. Two photodetectors (PDs) were utilized to measure the power of the transmitted light. Similar to our prior measurements on GO films,[63] BP,[22] BiOBr nanoflakes,[64] PdSe$_2$,[65] and CH$_3$NH$_3$PbI$_3$ perovskite nanosheets,[66] the Z-scan setup was calibrated to achieve a high accuracy before our Z-scan measurements.

In the open-aperture (OA) measurement, all transmitted light passing through the sample was collected by PD1 in **Figure 3a**, and the observed variation in the optical transmittance was induced by the nonlinear optical absorption of the sample. To determine the nonlinear absorption coefficient ($\beta$) of the MXene film, the measured OA results were fit with[64, 65]

$$T_{OA}(z) \simeq 1 - \frac{1}{2\sqrt{2}} \frac{\beta I_0 L_{eff}}{1 + z/z_0}, \qquad (1)$$

where $T_{OA}(z)$ is the normalized optical transmittance of the OA measurement, $I_0$ is the irradiance intensity at the focus, $z$ and $z_0$ are the sample position relative to the focus and the Rayleigh length of the laser beam, respectively, and $L_{eff} = (1 - e^{-\alpha_0 L})/\alpha_0$ is the effective sample thickness, with $\alpha_0$ and $L$ denoting the linear absorption coefficient and the sample thickness, respectively.

**Figure 3b** shows images of the focused laser beam on the sample at different positions along the Z-axis, which were recorded by the CCD camera in **Figure 3a**. The incident laser intensity was ~53.29 GW/cm$^2$, and no visible damages or changes were observed in the sample as a result. A hazy outline of the beam was observed when the sample was initially out of focus, and the radius of the beam was dispersed throughout the boundaries of the image. As the sample approached the focal point at $z = 0$ mm, the beam radius decreased, and the center became more brilliant. The beam became a brilliant speckle with a diameter of ~1.6 μm when the sample was at the focus point. Subsequently, the beam spread as the sample moved away from the focal point.

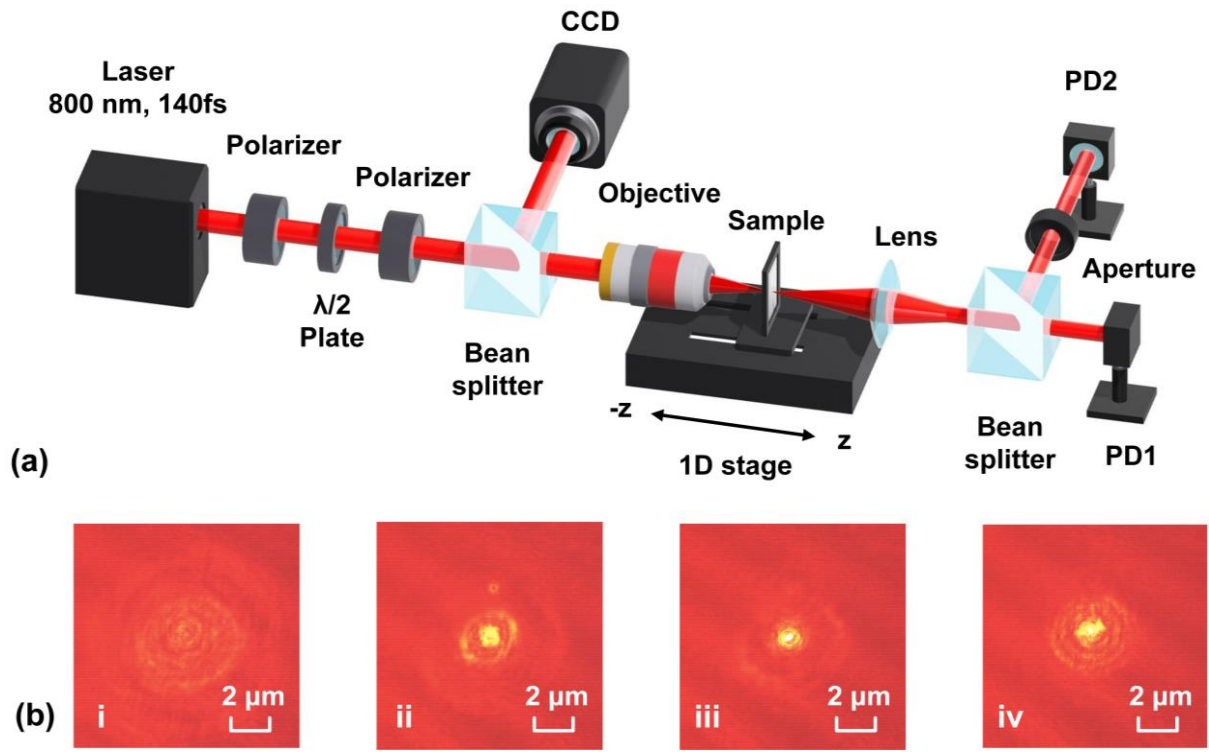

**Figure 3.** (a) Schematic illustration of the Z-scan experimental setup. CCD: charge-coupled device. PD: photo detector. (b) The focused laser beam on the sample at different positions along the Z-axis, where $z$ = (i) -0.04 mm, (ii) -0.015 mm, (iii) 0 mm, and (iv) 0.015 mm, respectively.

**Figures 4a–c** show the OA results for a 5-layer MXene film measured at varying incident laser intensities ranging from ~53.29 to ~222.04 GW/cm$^2$, respectively. In the OA curves, typical RSA, or optical limiting behavior was observed, with the transmission decreasing as the MXene sample approached the focal point. In addition, it was observed that the transmittance dip of the OA curve decreased as the incident laser intensity increased. In contrast, we did not observe any significant nonlinear absorption for an uncoated silica substrate and a silica substrate only coated with PDDA polymer, indicating that the MXene film was responsible for the observed nonlinear optical absorption. The modest deviation of experimental results from the standard symmetric OA curves can be attributed to the scattering from minor particles on the MXene samples as well as the irregularities and asymmetries in the input laser beam profile.

By fitting the measured OA results with **Eq. (1),** we derived the nonlinear absorption coefficient $\beta$ of the MXene films. **Figure 4d** shows the fit values of $\beta$ as a function of the irradiance laser intensity $I$. Five distinct locations on the sample were measured, and the data points represent their average values, with the error bars indicating the variations among these measurements. A large $\beta$ value of ~7.13 × 10$^2$ cm/GW is achieved at $I$ = ~53.29 GW/cm$^2$. It is also observed that absorption coefficient decreases as laser intensity increases. A similar phenomenon was previously reported in Ref. 26, which could be attributed to the alterations in the surface functional groups of the MXene films induced by the increased laser intensity.

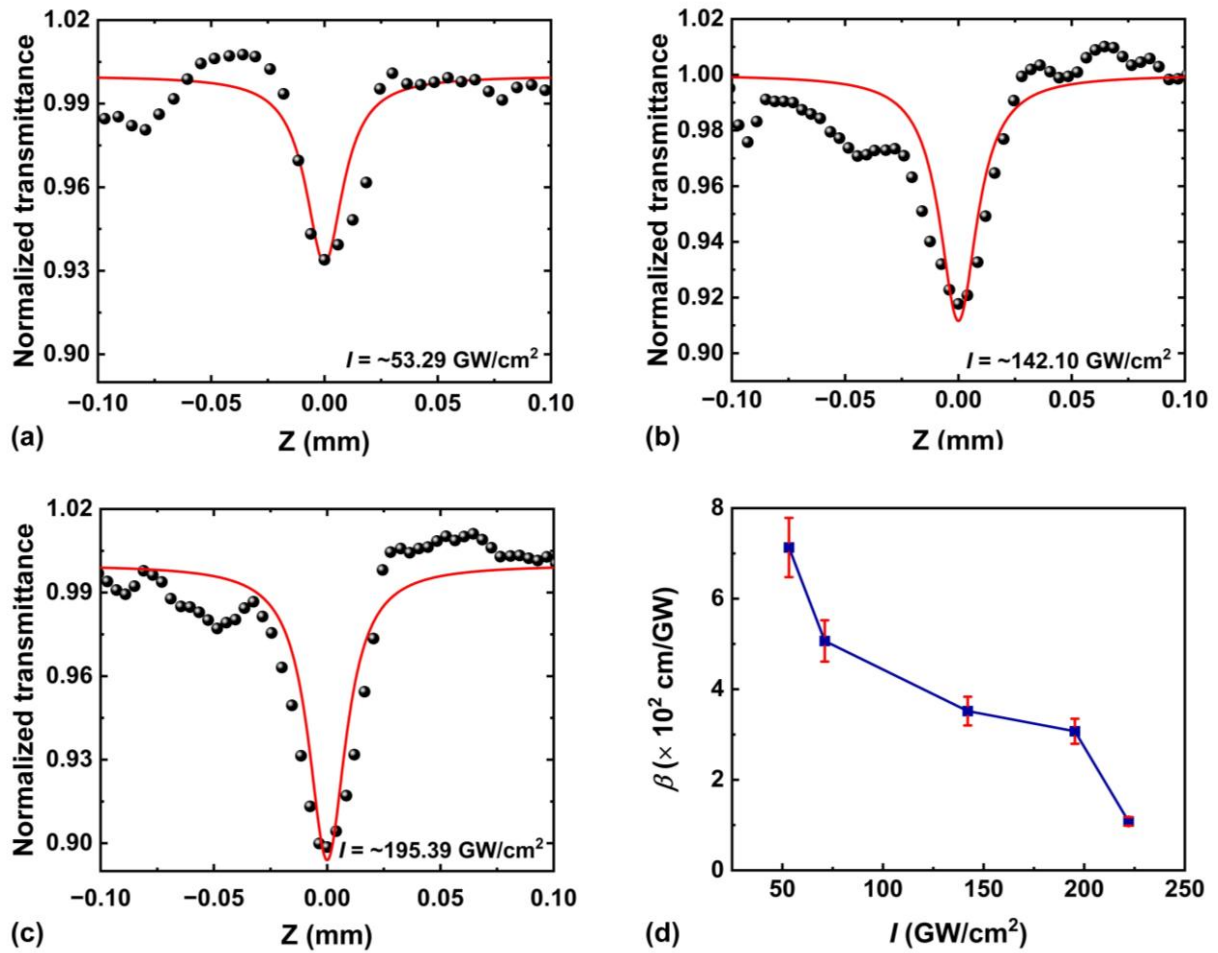

**Figure 4.** (a) – (c) OA Z-scan results of a 5-layer MXene film at different irradiance laser intensities of ~53.29, ~142.10, and ~195.39 GW/cm$^2$, respectively. (d) Nonlinear absorption coefficient $\beta$ of the MXene film versus irradiance laser intensity $I$.

**Figures 5a – e** depict the OA results for MXene films with different layer numbers $N$ = 5, 10, 15, 20, and 30. For comparison, all the samples had the same irradiance laser intensity of

~53.29 GW/cm². The fit $\beta$ as a function of $N$ is shown in **Figure 5f**. Similar to that in **Figure 4d**, we also measured five distinct locations for each sample. As $N$ increases from 5 to 30, the average value of the fit $\beta$ changes from ~7.13 × 10² cm/GW to ~-2.69 × 10² cm/GW. It is interesting to observe that the nonlinear absorption property of the MXene film exhibits significant layer dependence. For $N \leq 15$, the film displays typical RSA behavior, whereas SA behavior is observed for $N \geq 20$.

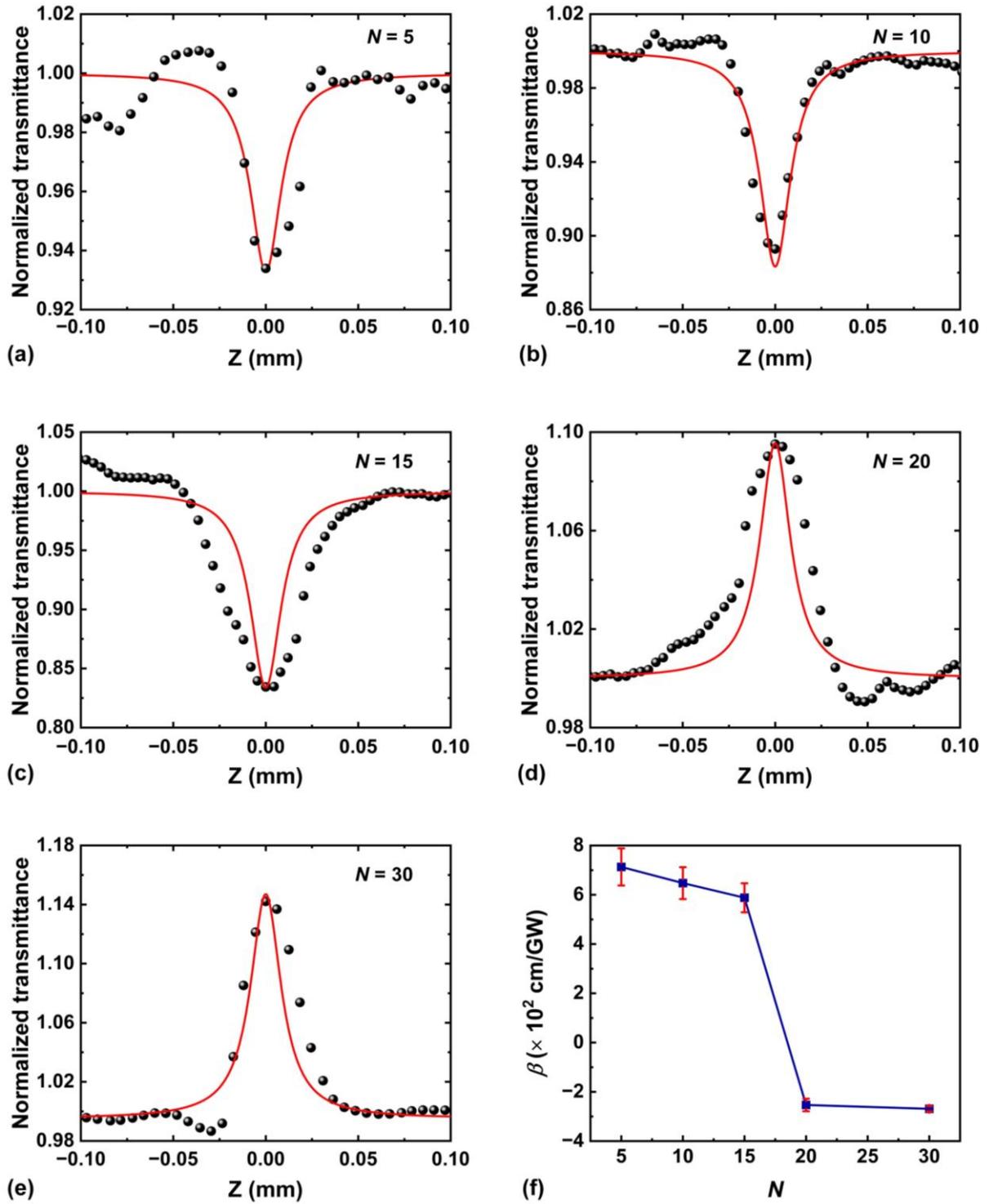

**Figure 5.** (a) – (e) Measured (data points) and fit (solid curves) OA results for MXene films with different layer numbers $N$ = 5, 10, 15, 20, and 30. (f) Fit the nonlinear absorption coefficient $\beta$ versus layer number $N$.

According to the UV−vis spectra in **Figure 2a**, the bandgap of the 5-layer MXene film is estimated to be ~1.0 eV, which is lower than the single photon energy of the incident light at ~800 nm, i.e., ~1.55 eV. As a result, the nonlinear absorption behavior when $N$ = 5 should be

SA, while our Z-scan measurement result in **Figure 5a** shows a positive $\beta$ that corresponds to the RSA. In principle, the RSA can be induced by various nonlinear optical effects such as nonlinear light scattering (NLS), two-photon absorption (TPA), and multiphoton absorption. These effects could co-exist in practical MXene films, thus resulting in complex and wavelength-dependent nonlinear optical absorption behavior. Considering that the NLS effect typically dominates in dispersion and solution-based materials with laser-induced microbubbles,[65] it should not be a significant factor for our prepared MXene films. Given the relatively low efficiency of multiphoton absorption, TPA is likely responsible for the occurrence of RSA in our case. For MXene films with lower thicknesses, apart from the prevailing TPA process, the electrons in the ground state $E_0$ can also transition to conduction band minimum (CBM) via one-photon absorption (1PA) after absorbing a single photon and emitting one phonon, as depicted in **Figure 6a**. One possible reason for the prevalence of the TPA over the 1PA could be its self-sufficiency, as opposed to 1PA, which relies on phonon assistance for its occurrence. Additionally, the ultrahigh peak power of femtosecond optical pulses makes TPA easily excitable. Previous study have suggested a relationship between TPA and the bandgap shape of MXene,[27] which can also explain the nonlinear optical absorption behavior for our MXene films. Specifically, the bandgap associated with single-photon energy could appear at the steeper edges of the cone structure. In contrast, at twice the photon energy, the bandgap exhibits a flatter shape, which makes TPA more favorable compared to 1PA.

As the film thickness increases (i.e., for $N \geq 20$), the nonlinear optical absorption observed in **Figures 5d** and **e** changes to SA, which can be attributed to thickness-dependent fluctuations in the bandgap. As depicted in **Figure 6b**, the bandgap of MXene decreases with the increase of layer number. This is also supported by the results in **Figures 2b** and **2c,** where a smaller bandgap of ~0.66 eV is for the 30-layer sample. As the bandgap reduces, the need for phonon assistance in 1PA diminishes due to a better alignment between the bandgap related to single-photon energy and the excited photon energy, which could facilitate a more accessible transition of ground state electrons to $E_1$ through 1PA. As a result, 1PA becomes more dominant in the nonlinear absorption behavior, leading to an SA behavior in the MXene films. We note that a similar nonlinear absorption behavior was also observed in 2D $WS_2$, $MoS_2$, $Bi_2S_3$ and $PtSe_2$[67,68] with different film thicknesses.

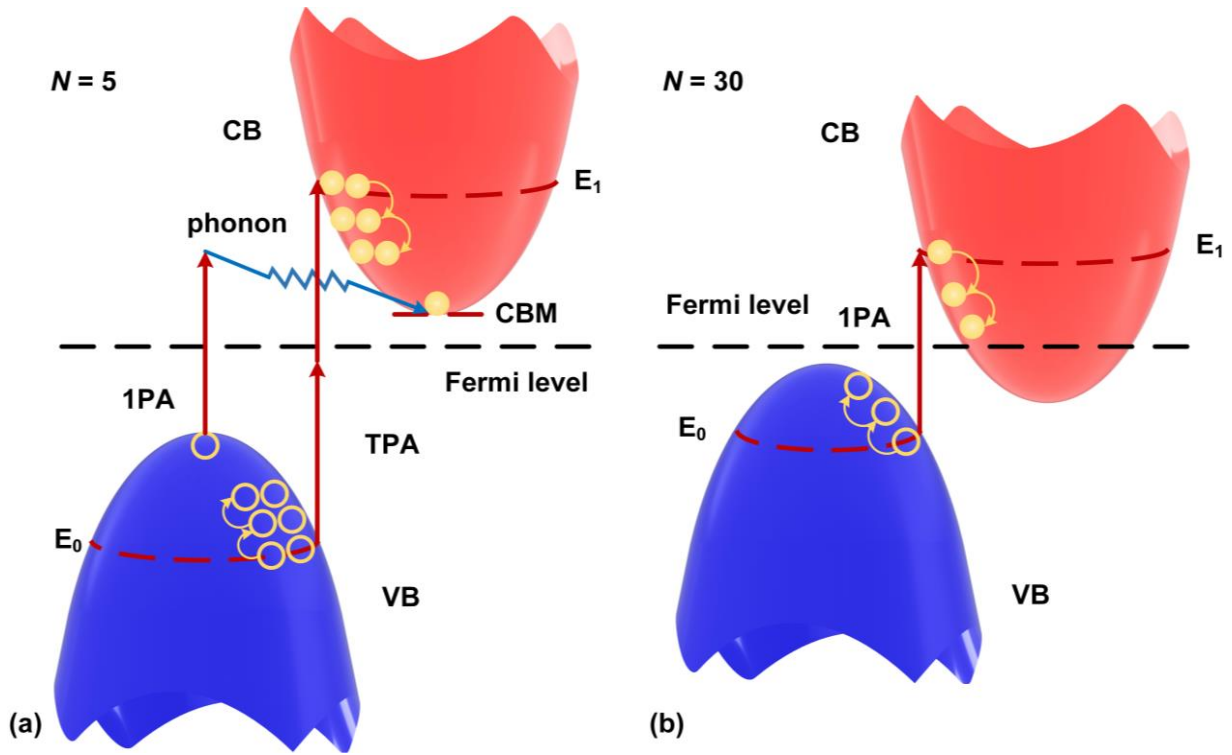

**Figure 6.** Schematic illustration of bandgap structures of (a) 5-layer and (b) 30-layer MXene films. 1PA: one-photon absorption. TPA: two-photon absorption. VB: valence band. CB: conduction band. CBM: conduction band minimum.

We also characterized the nonlinear refractive index $n_2$ of the MXene films using the closed aperture (CA) measurement. A small aperture was positioned before PD2 in **Figure 3** to collect only a portion of the on-axis transmitted light beam. However, we did not observe any significant z-scan curves, indicating that the $n_2$ is not high. Our observation here is consistent with the result in Ref. 26, where the measured $n_2$ value of MXene was ~$10^{-20}$ $m^2/W$ and was two orders of magnitude lower than that of silicon[1].

It is worth noting that damages to the MXene films were observed at very high laser powers, as evidenced by a sudden increase in the SA peaks, after which the OA results were no longer broadened by further increasing the laser power. We also find that samples with higher film thicknesses exhibit lower power endurability. For instance, the damages occurred at ~128.78 GW/cm$^2$ for the 20-layer film, while they occurred at ~71.05 GW/cm$^2$ for the 30-layer film.

In **Table I,** we compare our measured nonlinear absorption coefficient $\beta$ of MXene films with other 2D materials. For thin film including 5 layers of MXene, a positive $\beta$ on the order of $10^2$ cm/GW is obtained, which is comparable to BiOCl[69] and one orders of magnitude higher than BP[21] and PtSe$_2$.[67] On the other hand, a negative $\beta$ on the order of -$10^2$ cm/GW is obtained for thick film including 30 layers of MXene. This is significantly greater than MoS$_2$[70] and Ni-MOF.[71] All of these reflect the superior nonlinear optical properties of the MXene films. In addition, the MXene films' capability to exhibit a change in the sign of $\beta$ for different film thicknesses greatly expands the range of potential nonlinear optical devices for diverse applications beyond those of a material with a single sign of $\beta$. The strong and layer-dependent nonlinear absorption, together with the facile film coating method with large-area coating capability and precise control of the film thickness, could potentially underpin a variety of nonlinear optical devices incorporating 2D MXene films, as we have demonstrated for the GO films fabricated via a similar method. The significant and layer-dependent nonlinear absorption of the 2D MXene films, combined with the facile film coating process allowing large-area coverage and precise film thickness control, opens up new possibilities for implementing nonlinear optical devices with high performance and new features. As demonstrated previously with 2D GO films fabricated using a similar approach,[3, 17, 18, 33, 54, 76-123] the integration of MXene films into photonic devices shows promising potential for a wide range of nonlinear optical applications[124-149].

**Table I. Comparison of nonlinear absorption coefficient $\beta$ for various 2D materials.**

| Material | Laser parameter | Film thickness | $\beta$(cm/GW) | Ref. |
|---|---|---|---|---|
| Graphene | 1550 nm, 100 fs | 5−7 layers | $9 \times 10^3$ | 72 |
| GO | 800 nm, 100 fs | ~2 μm | $4 \times 10^4$ | 63 |
| $MoS_2$ | 1064 nm, 25 ps | ~25 μm | $-3.8 \pm 0.59$ | 70 |
| $WS_2$ | 1040 nm, 340 fs | ~57.9 nm | $(1.81 \pm 0.08) \times 10^3$ | 73 |
| $WSe_2$ | 1040 nm, 340 fs | 25.1 nm | $2.14 \times 10^3$ | 73 |
| $PdSe_2$ | 800 nm, 140 fs | ~8 nm | $(3.26 \pm 0.19) \times 10^3$ | 65 |
| $PtSe_2$ | 1030 nm, 340 fs | 4 layers | $2.96 \pm 0.05$ | 67 |
| BP | 800 nm, 100 fs | 30 – 60 nm | $45 \pm 2$ | 21 |
| h-BN | 400 nm, 150 fs | ~2.5 nm | $5 \times 10^4$ | 74 |
| $Bi_2Te_3$ | 1056 nm, 100 fs | 24 – 25 nm | $2.29 \times 10^6$ | 75 |
| BiOCl | 800 nm, 100 fs | 20 – 140 nm | $4.25 \times 10^2$ | 69 |
| BiOBr | 800 nm, 140 fs | 140 | $1.869 \times 10^4$ | 64 |
| Ni-MOF | 800 nm, 95 ± 10 fs | ~4.2 nm | $-3 \times 10^{-2}$ | 71 |
| MXene | 800 nm, 140 fs | 5 layers | $7.13 \times 10^2$ | This work |
| MXene | 800 nm, 140 fs | 30 layers | $-2.69 \times 10^2$ | This work |

## 4. CONCLUSION

In summary, we prepare layered MXene films via a solution-based method that yields transfer-free and layer-by-layer film coating and investigate their layer-dependent nonlinear absorption properties. The results of our Z-scan measurements at 800 nm show that the MXene films exhibit a strong layer-dependent nonlinear absorption behavior. As the layer number increases from 5 to 30, the films transition from reversible saturable absorption (RSA) to saturable absorption (SA), accompanied by a noteworthy variation in the nonlinear absorption coefficient $\beta$, ranging from ~$7.13 \times 10^2$ cm/GW to ~$-2.69 \times 10^2$ cm/GW. In addition, we characterize the nonlinear absorption of the MXene films at varying incident laser intensities, finding that the nonlinear absorption coefficient $\beta$ decreases as the laser intensity increases. These results reveal the interesting layer-dependent nonlinear optical properties of 2D MXene films, which will facilitate the implementation of high-performance MXene-based nonlinear photonic devices.

**Notes**

The authors declare no competing interest.